\documentclass[twocolumn]{aastex61} 
\usepackage{graphicx}				
\usepackage{amssymb}
\usepackage{amsmath}

\begin{document}

\title{Populating the low-mass end of the $M_{\rm BH}-\sigma_{\ast}$ relation }
\author{Vivienne F. Baldassare}
\altaffiliation{Einstein Fellow}
\affiliation{Yale University \\
Department of Astronomy \\
52 Hillhouse Avenue \\
New Haven, CT 06511, USA}

\author{Claire Dickey}
\affiliation{Yale University \\
Department of Astronomy \\
52 Hillhouse Avenue \\
New Haven, CT 06511, USA}

\author{Marla Geha}
\affiliation{Yale University \\
Department of Astronomy \\
52 Hillhouse Avenue \\
New Haven, CT 06511, USA}

\author{Amy E. Reines}
\affiliation{eXtreme Gravity Institute \\
Department of Physics \\
Montana State University\\
Bozeman, MT 59717 USA}

\correspondingauthor{Vivienne F. Baldassare} \email{vivienne.baldassare@yale.edu}

\received{28 May 2020}
\accepted{26 June 2020}
\submitjournal{ApJL}

\begin{abstract}

We present high resolution spectroscopy taken with the Keck Echellete Spectrograph and Imager to measure stellar velocity dispersions for eight active dwarf galaxies ($M_{\ast}<3\times10^{9}~M_{\odot}$) with virial black hole masses. We double the number of systems in this stellar mass regime with measurements of both black hole mass ($M_{\rm BH}$) and stellar velocity dispersion ($\sigma_{\ast}$), and place them on the $M_{\rm BH}-\sigma_{\ast}$ relation. The tight relation between $M_{\rm BH}$ and $\sigma_{\ast}$ for higher mass galaxies is a strong piece of evidence for the co-evolution of BHs and their host galaxies, but it has been unclear whether this relation holds in the dwarf galaxy regime. 
Our sample is in good agreement with the extrapolation of the $M_{\rm BH}-\sigma_{\ast}$ relation to low BH/galaxy masses, suggesting that the processes which produce $M_{\rm BH}-\sigma_{\ast}$ can also operate in dwarf galaxies. These results provide important constraints for massive black hole seed formation models and models exploring the impact of AGN feedback in dwarf galaxies.

\end{abstract}

\section{Introduction}

Central massive black holes (BHs) exist in virtually all galaxies with stellar masses $M_{\ast}\gtrsim10^{10}\;M_{\odot}$ \citep{1998AJ....115.2285M}. There are scaling relations between the mass of the central BH and properties of the host galaxy \citep{Kormendy:2013ve}, such as galaxy stellar mass \citep{2015ApJ...813...82R} , the mass/luminosity of the stellar bulge \citep{2014ApJ...780...70L, 2019ApJ...887..245S}, and the velocity dispersion of stars in the bulge \citep{2000ApJ...539L...9F, 2000ApJ...539L..13G,2009ApJ...698..198G}. These scaling relations are key to our understanding of how BH growth relates to the evolution of galaxies.

The most well studied of these scaling relations is between stellar velocity dispersion and BH mass, also known as the $M_{\rm BH}-\sigma_{\ast}$ relation.  There is a large body of work exploring the $M_{\rm BH}-\sigma_{\ast}$ relation for relatively high-mass galaxies (e.g., \citealt{2009ApJ...698..198G, Kormendy:2013ve, 2013ApJ...764..184M, 2013ApJ...772...49W, 2016ApJ...831..134V, 2018MNRAS.477.3030K}), and its existence points towards feedback between the growth of the BH and its host galaxy.

The low-mass end of the $M_{\rm BH}-\sigma_{\ast}$ relation is of particular importance as it is predicted to give insight into the formation mechanisms of BH seeds at high redshift, as well as into the efficiency of BH growth in small galaxies \citep{2019arXiv191109678G}. In particular, different BH seed formation mechanisms (i.e., Pop III stars versus direct collapse) may be reflected in the the slope and scatter of the low-mass end of $M_{\rm BH}-\sigma_{\ast}$ \citep{2009MNRAS.400.1911V}. Lighter Pop III seeds are predicted to produce a present-day population of under-massive BHs, while heavier direct collapse seeds would result in a flattening of  $M_{\rm BH}-\sigma_{\ast}$ around BH masses of $\sim10^{5}\;M_{\odot}$.  BH fueling may also impact where low-mass BHs/galaxies fall on the relation \citep{2018MNRAS.481.3278R}. \cite{2018ApJ...864L...6P} predicts that BH accretion should be bimodal, with BHs $\lesssim 10^{5}\;M_{\odot}$ accreting inefficiently, resulting in a population of low-mass BHs that fall beneath the present-day $M_{\rm BH}-\sigma_{\ast}$ relation.

Unfortunately, it has been difficult to probe $M_{\rm BH}-\sigma_{\ast}$ in the dwarf galaxy regime. In addition to the observational difficulty of detecting low-mass BHs \citep{2016PASA...33...54R, 2018ApJ..868..152, 2019arXiv191006342B, 2020ApJ...888...36R}, measuring $\sigma_{\ast}$ in these systems requires high spectral resolution observations of faint galaxies. 
In the last several years, large-scale optical spectroscopic surveys have helped with the former; there has been a substantial increase in the number of known active galactic nuclei (AGN) in low-mass galaxies, which can now be used to probe scaling relations at the low-mass end \citep{Reines:2013fj, 2014AJ....148..136M, 2015ApJ...809L..14B, 2015MNRAS.454.3722S}.

Motivated by the power of scaling relations for exploring BH formation and fueling, we obtained new stellar velocity dispersion measurements for eight dwarf galaxies  ($M_{\ast}<3\times10^{9}\;M_{\ast}$) with low-mass BHs and explore the low-mass end of $M_{\rm BH}-\sigma_{\ast}$. With our observations, we double the number of dwarf galaxies on $M_{\rm BH}-\sigma_{\ast}$.  Section 2 describes our sample and observations; Section 3 discusses our velocity dispersion and BH mass measurements; Section 4 presents the $M_{\rm BH}-\sigma_{\ast}$ relation including our low-mass systems and discusses implications. 

\section{Sample and Observations}

\begin{figure}
    \centering
    \includegraphics[width=0.5\textwidth]{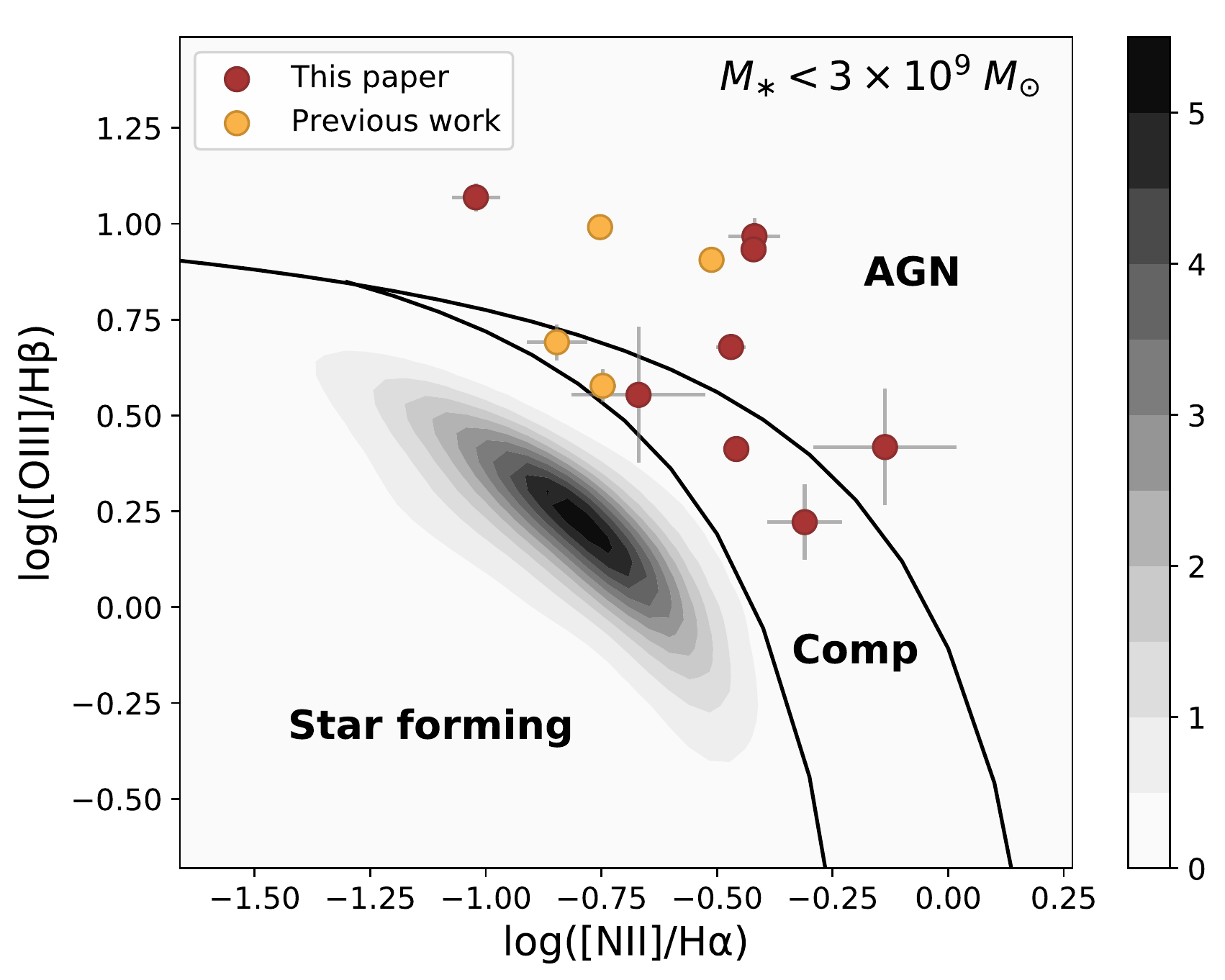}
    \caption{BPT diagram of the active dwarf galaxies we place on the $M_{\rm BH}-\sigma_{\ast}$ relation. The points include the 10 broad-line AGN from \cite{Reines:2013fj}, RGG 118, and Pox 52. Galaxies for which we have obtained new velocity dispersion measurements are shown as red circles; objects with existing measurements are shown in orange. The shaded contours show the positions of galaxies with $M_{\ast}<3\times10^{9}\;M_{\odot}$ from the NASA-Sloan Atlas (30654 galaxies). Line flux measurements for the active dwarf galaxies are from \cite{Reines:2013fj}; fluxes for the other systems are from the NASA-Sloan Atlas. }
    \label{fig:bpt}
\end{figure}

\begin{figure*}
\centering
\includegraphics[width=0.85\textwidth]{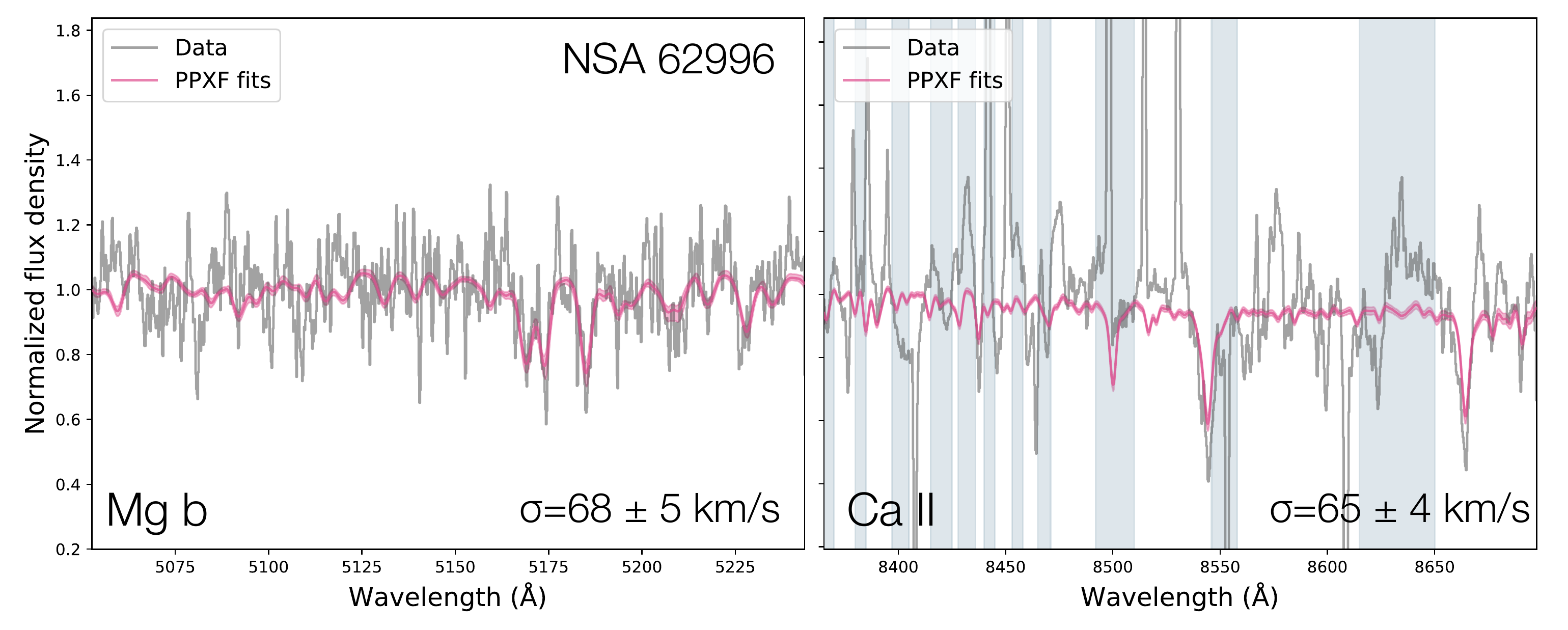}
\caption{ESI spectrum and pPXF fits for NSA 62996. In each panel, the data is shown in gray and the best-fit pPXF models are shown in pink. The solid pink line is the mean pPXF fit, and the shaded pink region encompasses all 1000 fits. The left panel shows the region surrounding the Mg $\rm{I b}$ triplet, and the right panel shows the region surrounding the Ca II triplet. The locations of sky lines are masked in fitting and marked by shaded blue regions in the figure. The velocity dispersion and uncertainties are given in the bottom right of each panel. Spectrum is smoothed with a box size of 3 pixels for plotting. Additional spectra shown in Figures~\ref{additional_sigs} and \ref{additional_sigs2}.}
\label{nsa52675_panels}
\end{figure*}

In this work, we measure stellar velocity dispersions for eight nearby ($z<0.055$) active dwarf galaxies with optical spectroscopic signatures of AGN activity. These systems are drawn from the sample of \cite{Reines:2013fj} dwarf galaxies with both broad and narrow optical emission line signatures of AGN activity (using the BPT diagram; \citealt{1981PASP...93....5B, 2006MNRAS.372..961K}). They were originally selected from the NASA-Sloan Atlas\footnote{nsatlas.org}, a catalog of local galaxies with SDSS imaging and spectroscopy and derived quantities such as stellar mass. 
\cite{Reines:2013fj} identified 10 systems meeting the above criteria; two of these have velocity dispersions measured in previous works. The observations presented here complete the velocity dispersion measurements for all of the \cite{Reines:2013fj} active dwarf galaxies with BH mass estimates. Figure~\ref{fig:bpt} shows where the active dwarf galaxy sample falls on the BPT diagram. Stellar masses range from $8\times10^{8} - 3\times10^{9}\;M_{\odot}$.

In addition to the eight new systems presented here, we include seven dwarf galaxies with velocity dispersions and BH masses measured in the literature. These include NGC 4395 \citep{1989ApJ...342L..11F, 2003ApJ...588L..13F}, Pox 52 \citep{2004ApJ...607...90B, 2008ApJ...686..892T}, RGG 118 \citep{2015ApJ...809L..14B, 2017ApJ...850..196B}, RGG 119 \citep{2016ApJ...829...57}, M 32 \citep{2010MNRAS.401.1770V}, and NGC 5206 and NGC 205 \citep{2018ApJ...858..118N, 2019ApJ...872..104N}.

Observations were taken with the Keck II Echellette Spectrograph and Imager (ESI; \citealt{2002PASP..114..851S}) on 10 March 2018 and 4 September 2019. Observations were made with the 0.75$''$ x 20$''$ slit, which gives an instrumental resolution of 23 $\mathrm{km \ s^{-1}}$ across the wavelength range 3900-11000 \AA. The dispersion ranges from $0.16 \rm{\AA / pixel}$ in the blue to $0.30 \rm{\AA / pixel}$ in the red, giving a constant velocity dispersion of 11.5 $\rm{km\;s^{-1}\; pixel^{-1}}$. Total exposure times ranged from 1200-2700 seconds, and were split into three exposures to facilitate cosmic ray removal. The per-pixel signal-to-noise ratios of the spectra range from $\sim3-20$, with a median signal-to-noise ratio of 8.

We used \texttt{XIDL}\footnote{ https://www2.keck.hawaii.edu/inst/esi/ESIRedux/index.html} \citep{prochaska2003} for the initial reduction of the science frames. Separate wavelength solutions were derived for each night of observations from a combination of CuAr and HgNe+Xe arcs. Cosmic rays were removed from each frame with \texttt{LACosmic} \citep{vandokkum2001}. Individual exposures were then median combined and rectified so that the spatial scale of the pixels are the same across all orders. The background sky was modelled with a bspline fit to the outer 2$''$ on each side of the slit.

1-D spectra were optimally extracted \citep{1986PASP...98..609H} using a width of 6 pixels, which corresponds to an extraction width of $\sim1''$. At the distances of our sample, this corresponds to $0.5-1$ kpc.

\section{Analysis}

\subsection{Velocity dispersions}

We use the Penalized Pixel Fitting software (pPXF; \citealt{2004PASP..116..138C, cappellari2017}) to measure stellar velocity dispersions. pPXF fits the absorption line spectra of galaxies using a library of stellar spectra and extracts galaxy stellar kinematics. We use a library of 9 stars ranging in spectral type from F to M. They were observed with ESI using the same slit width as the galaxy observations. The best fits include a combination of several stellar spectra. 
We fit the kinematics in two regions: one surrounding the Mg $\rm{I b}$ triplet at $\sim5160-5190 {\rm \AA}$ and another surrounding the Ca II triplet at $\sim8490-8670 {\rm \AA}$. The uncertainty in the velocity dispersion is computed with a Monte Carlo bootstrap method \citep{2009ApJ...692.1464G}. We add noise to the 1-D spectrum, then recalculate the velocity dispersion for 1000 noise realizations. The final value is taken to be the mean recovered velocity dispersion, and the uncertainty is taken to be the square root of the variance relative to the mean.

\begin{deluxetable*}{cccccccc}
\tablecaption{Dwarf galaxies with stellar velocity dispersions and black hole masses \label{veldisp_tab}}
\tablecolumns{8}
\tablenum{1}
\tablewidth{0pt}
\tablehead{
\colhead{Name} &
\colhead{R13 ID} &
\colhead{Redshift} & 
\colhead{Morphology} & 
\colhead{$\log_{10}(M_{\ast}/M_{\odot})$} &
\colhead{$\log_{10}(M_{\rm BH}/M_{\odot})$} &
\colhead{$\sigma_{\ast}$}  & 
\colhead{Reference} \\
\colhead{} &
\colhead{} &
\colhead{} &
\colhead{} &
\colhead{} &
\colhead{} &
\colhead{[km/s]} 
}
\startdata
NSA 62996 & 1 & 0.0459 & S0 & 9.45 & 5.80 &  $66 \pm 3 $  & This paper  \\
NSA 10779 & 9 & 0.0466 & dE & 9.30 & 5.44 &  $34 \pm 6 $ $^{\rm a}$& This paper \\
NSA 125318  & 11 & 0.0327 & S0 & 9.24 & 5.00 & $41 \pm 6$ & This paper \\
NSA 52675 & 20 & 0.0144 & S0 & 9.29 & 6.10 & $53 \pm 5 $ & This paper   \\  
NSA 15235  & 32 & 0.0299 & Spiral & 9.30 & 5.29  & $42 \pm 14$ $^{\rm a}$ & This paper \\
NSA 47066 & 48 & 0.0410 & Spiral & 9.12 & 5.42 &  $32 \pm 5$ $^{\rm a}$ & This paper\\
NSA 18913 & 123 & 0.0395 & Disk & 8.96 &  5.18 &  $33\pm14$ $^{\rm a}$ & This paper \\
NSA 99052 & 127 & 0.0317 & Disk & 9.36 & 5.21 & $52 \pm 5$ & This paper\\
\hline
NGC 4395  & 21 & 0.0011 & Spiral & 9.10 & 5.0 &  $<30$ & \cite{2003ApJ...588L..13F} \\
NSA 166155 & 118 & 0.0243 & Spiral & 9.34  & 4.7 & $28 \pm 11$ & \cite{2015ApJ...809L..14B}   \\
NSA 79874 & 119 & 0.0384 & S0 & 9.36 &  5.46  &  $28\pm7$ & \cite{2016ApJ...829...57} \\
Pox 52 & N/A &  0.022 & dE & 9.08  & 5.2 & $36\pm5$ & \cite{2004ApJ...607...90B} \\
NGC 205 & N/A & -0.0008 & dE & 8.99 & 3.8$^{\rm b}$ & $40 \pm 5$  & \cite{2018ApJ...858..118N, 2019ApJ...872..104N}  \\
NGC 5206 & N/A  & 0.002 & dE  & 9.38 & 5.8$^{\rm b}$ & $35 \pm 1$ & \cite{2018ApJ...858..118N, 2019ApJ...872..104N}  \\
M 32 & N/A & -0.0007 & dE &  9.0 & 6.4$^{\rm b}$ & $77\pm 3$ & \cite{2010MNRAS.401.1770V} \\ 
\enddata
\tablecomments{Black hole masses and velocity dispersion measurements. $^{\rm a}$ indicates that the velocity dispersion measurement uses Mg Ib only.  $^{\rm b}$ indicates a dynamical BH detection and mass estimate. Galaxy stellar masses and BH masses for objects in the \cite{Reines:2013fj} sample are taken from \cite{2015ApJ...813...82R}. Morphologies are taken from \cite{2019ApJ...887..245S} and the NASA/IPAC Extragalactic Database. }
\end{deluxetable*}

At the redshifts of our sample, the Ca II triplet falls at the end of the ESI spectrum in a region with substantial contamination from sky lines. While we measure Mg $\rm{I b}$ velocity dispersions for all eight objects, it is only possible to obtain Ca II measurements for four. When we can fit both regions, the estimates are consistent with one another and we adopt the mean as the final velocity dispersion measurement. For the rest, we adopt the Mg $\rm{I b}$ measurements. Figure~\ref{nsa52675_panels} shows the ESI spectra and pPXF fits to the Mg $\rm{I b}$ and Ca II regions for NSA 62996. Table~\ref{veldisp_tab} gives the velocity dispersion measurements for the systems analyzed here along with measurements for seven dwarf galaxies reported in previous works. 

The values reported in Table~\ref{veldisp_tab} correspond to those measured in the 1$''$ extracted spectrum. This is well-matched to the bulge properties found for this sample by \cite{2019ApJ...887..245S} using HST observations; their median bulge diameter is 0.6 kpc. Here, the term ``bulge" is used to refer to the inner S{\'e}rsic component, though most have S{\'e}rsic indices that are below that of a classical bulge. We also measure the stellar kinematics in spectra  extracted between 0.5 and 1.5$''$ on either side of the central arcsecond; we find that these values are consistent with flat velocity dispersion profiles. This is in good agreement with the velocity dispersion profiles measured for more nearby dwarf elliptical galaxies \citep{2006AJ....131..332G, 2010ApJ...711..361G, 2014ApJS..215...17T}.

\subsection{Black hole masses}

BH masses for the eight systems analyzed in this work are taken from \cite{2015ApJ...813...82R}. The BH masses are single-epoch spectroscopic masses computed using the broad H$\alpha$ emission line. Under the assumption that gas in the broad line emitting region is virialized, one can estimate the BH mass as $M_{\rm BH} = f \frac{v^{2} R_{\rm BLR}}{G}$, where $v$ is the characteristic velocity of gas in the broad line region, $R_{\rm BLR}$ is the distance to the broad line region, and $f$ is a factor intended to take into account the unknown broad line region geometry. The full-width half maximum of the broad H$\alpha$ line is used to estimate the characteristic velocity, and the luminosity of broad H$\alpha$ is a proxy for the distance to the broad line region \citep{2005ApJ...630..122G, 2009ApJ...705..199B, 2013ApJ...767..149B}. BH masses estimated using this technique have systematic uncertainties of $\sim0.4$ dex. We do not recompute BH masses using the ESI data, but do confirm that the broad H$\alpha$ line widths are consistent between the SDSS and ESI data.

\begin{figure*}
\centering
\includegraphics[width=0.85\textwidth]{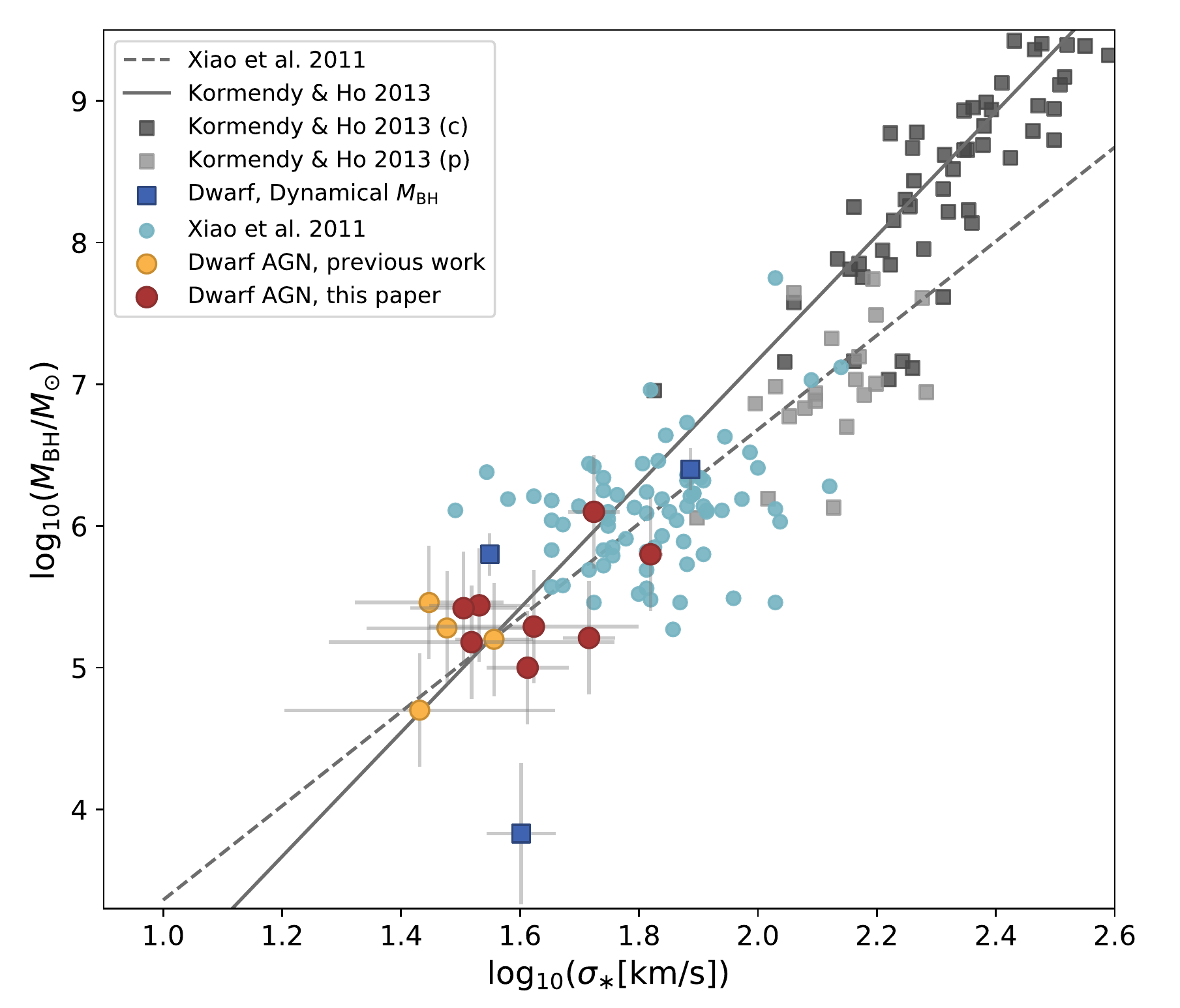}
\caption{Black hole mass versus stellar velocity dispersion. The gray squares are from the compilation of \cite{Kormendy:2013ve}; dark gray squares show galaxies with classical bulges and light gray squares show galaxies with pseudo-bulges. Light blue circles show data for low-mass AGNs from \cite{2011ApJ...739...28X}. Active dwarf galaxies with ESI stellar velocity dispersion measurements measured here are shown as red circles; active dwarf galaxies with existing data are shown in orange. Dwarf galaxies with dynamical BH mass estimates are shown as dark blue squares. We also show the fits to $M_{\rm BH}-\sigma_{\ast}$ from \cite{2011ApJ...739...28X} (dashed line) and \cite{Kormendy:2013ve} (solid line). }
\label{Msig}
\end{figure*}

\section{The $M_{\rm BH}-\sigma_{\ast}$ relation with dwarf galaxies}

Figure~\ref{Msig} shows the locations of the eight active dwarf galaxies analyzed here on the $M_{\rm BH}-\sigma_{\ast}$ relation, as well as the positions of the seven dwarf galaxies from the literature. 
We also include the sample of low-mass AGN ($M_{\rm BH}\lesssim2\times10^{6}\;M_{\odot}$) with measured stellar velocity dispersions from \cite{2011ApJ...739...28X} and galaxies with dynamical BH mass measurements from the \cite{Kormendy:2013ve} compilation. Most of the dwarf galaxies in the active sample are in good agreement with both the the $M_{\rm BH}-\sigma_{\ast}$ relation found by \cite{Kormendy:2013ve} using dynamical BH masses, and the relation found by \cite{2011ApJ...739...28X} for broad-line AGNs in the \cite{Greene:2007kx} sample. 

\begin{figure*}
\centering 
\includegraphics[width=0.93\textwidth]{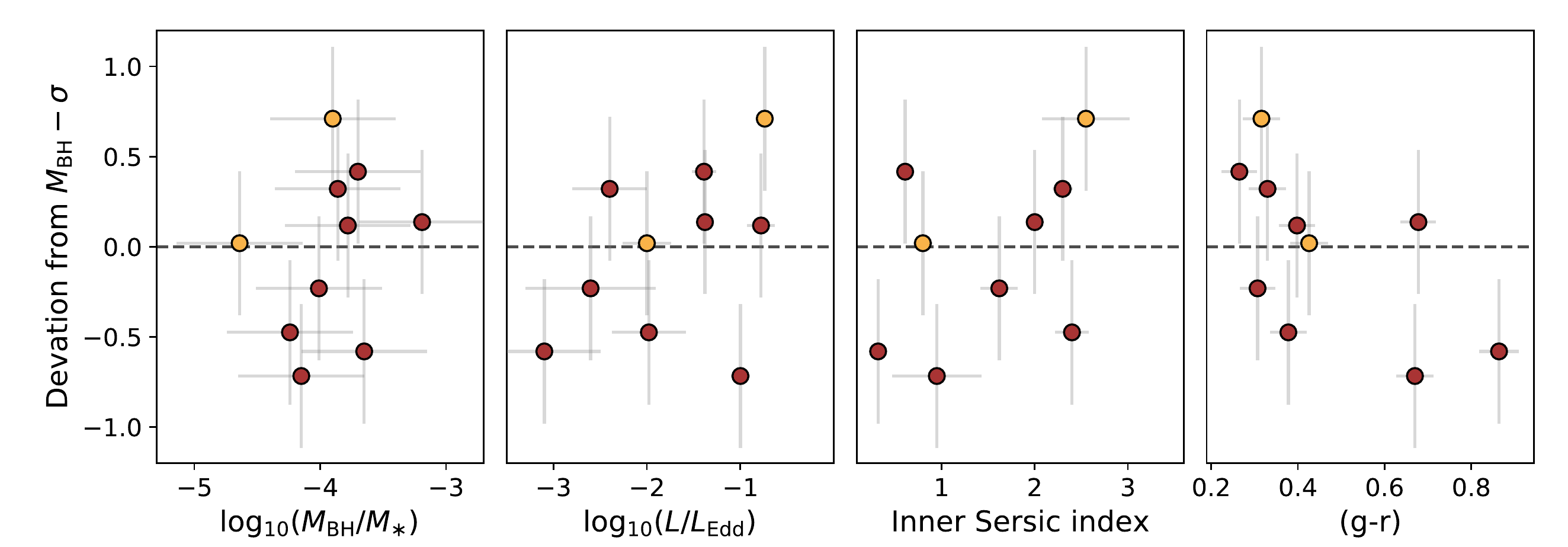}
\caption{Deviation from $M_{\rm BH}-\sigma_{\ast}$ relation as a function of galaxy/BH properties. Color-coding is the same as in Figures~\ref{fig:bpt} and \ref{Msig}. Deviation from $M_{\rm BH}-\sigma_{\ast}$ is defined as $\log_{10}(M_{\rm BH, measured}/M_{\rm BH, expected})$, where $M_{\rm BH, expected}$ is the BH mass predicted from the \cite{Kormendy:2013ve} $M_{\rm BH}-\sigma_{\ast}$ relation. From left to right, the deviation is plotted against the BH mass-to-stellar mass ratio, Eddington fraction, S{\'e}rsic index of the inner component, and dynamical mass-to-light ratio. }
\label{devs}
\end{figure*}

It is interesting to consider galaxy-scale properties which may influence BH growth and thus result in a galaxy falling above or below the $M_{\rm BH}-\sigma_{\ast}$ relation. In Figure~\ref{devs} we plot the deviation from the \cite{Kormendy:2013ve} relation versus galaxy properties to search for properties which may impact BH growth and produce over/under-massive BHs relative to $M_{\rm BH}-\sigma_{\ast}$. We consider the BH mass-to-stellar mass ratio, Eddington fraction, inner S{\'e}rsic index, and g-r color. Stellar masses are taken from \cite{2015ApJ...813...82R}. Eddington fractions are computed using X-ray luminosities reported in \cite{2012ApJ...761...73D} and \cite{2017ApJ...836...20B} and a bolometric correction of 10 \citep{Marconi:2003fk}. The bulge S{\'e}rsic indexes were determined using \textit{Hubble Space Telescope} imaging by \cite{2011ApJ...742...68J} and \cite{2019ApJ...887..245S}. The (g-r) color is computed using the NASA-Sloan Atlas Galactic-extinction corrected fluxes.

According to Spearman rank correlation coefficients, we find no significant correlations between the above properties and the deviation from the $M_{\rm BH}-\sigma_{\ast}$ relation, though we are limited by a relatively small sample size. The median ratio of BH to galaxy stellar mass is $\sim10^{-4}$, roughly an order of magnitude lower than more massive elliptical galaxies \citep{2015ApJ...813...82R}. The lack of a correlation between $M_{\rm BH}-\sigma_{\ast}$ deviation and Eddington fraction suggests that the current Eddington fraction does not reflect sustained BH growth over a long period of time. Similarly, the lack of correlation between bulge S{\'e}rsic index and $M_{\rm BH}-\sigma_{\ast}$ deviation suggests BH growth in dwarf galaxies is not necessarily related to the assembly of a bulge. The lack of correlation between galaxy color and $M_{\rm BH}-\sigma_{\ast}$ offset could reflect no difference in galaxy stellar populations; a more careful analysis involving a decomposition of galaxy light profiles and removal of the AGN contribution is necessary to fully explore this.

There are several recent theoretical works which suggest that the growth of lower-mass BH seeds may be inefficient (e.g., \citealt{2016arXiv160509394H, 2017MNRAS.472L.109A}), and that this should be reflected in the $M_{\rm BH}-\sigma_{\ast}$ relation. \cite{2018ApJ...864L...6P} use a semi-analytical model to track the growth of BH seeds and predict the present-day low-mass end of the $M_{\rm BH}-\sigma_{\ast}$ relation. Their model includes high and low-mass BH seeds formed via direct collapse and Pop III models, respectively. They find that, in general, systems deviate from $M_{\rm BH}-\sigma_{\ast}$ at low BH masses. This is because high mass BH seeds ($\gtrsim10^{5}\;M_{\odot}$) can more easily accrete in a ``high-efficiency regime"- a region in the 2-D parameter space of BH mass and gas number density that corresponds to sustained super-Eddington accretion \citep{2017ApJ...850L..42P}. 
On the other hand, low-mass seeds almost never accrete in the high-efficiency regime. By present day, \cite{2018ApJ...864L...6P}  predict that the $M_{\rm BH}-\sigma_{\ast}$ relation should become steeper at low BH masses, with observational biases producing any observed flattening. They predict this down-turn should occur at $\sigma_{\ast}\approx65{\rm km\;s^{-1}}$. 

The above implies that we find systems which fall on $M_{\rm BH}-\sigma_{\ast}$ because those are the BHs that are possible to find, while a population of under-massive BHs remains hidden. Our results -- a population of low-mass systems which largely agree with $M_{\rm BH}-\sigma_{\ast}$, as well as some which scatter below it -- are consistent with such a scenario. Finding BHs with masses $\lesssim10^{5}\;M_{\odot}$ and which fall below $M_{\rm BH}-\sigma_{\ast}$ also disfavors models which include only heavy BH seeds formed via direct collapse (i.e., those that predict BH seeds which form with $M_{\rm BH}=10^{5}\;M_{\odot}$). It is also important to note that the SDSS spectroscopy is only sensitive enough to detect broad H$\alpha$ emission from a $\sim10^{4} M_{\odot}$ BH accreting at its Eddington ratio \citep{Reines:2013fj}, so we are only sensitive to the most massive BHs in dwarf galaxies.

Our results also have interesting implications for AGN feedback in dwarf galaxies.  While dwarf galaxies are predominantly influenced by environment-triggered feedback \citep{2012ApJ...757...85G},  recent observational studies have found  tantalizing evidence for AGN-driven quenching in dwarf galaxies \citep{2018MNRAS.476..979P, 2019ApJ...884..180D}.  This is supported by recent simulations; \cite{2019arXiv191206646S} study AGN feedback in dwarf galaxies in Romulus25 and find that BHs can quench galaxies of similar mass to those considered in this work.  However, the exact mechanisms by which BHs influence dwarf galaxies remains uncertain, with some zoom-in simulations finding that AGN struggle to regulate global star formation rates \citep{2019MNRAS.484.2047K}.

Increasing the sample of dwarf galaxies on the $\mathrm{M_{BH}-\sigma_*}$ relation is critical for understanding the frequency and extent to which BHs influence galaxy evolution at low stellar masses. Our new sample of dwarf galaxies does not produce excessive scatter from the relation, suggesting that BH-galaxy co-evolution is still occurring at $\mathrm{M_* \sim 10^9 \ M_\odot}$. In high-mass galaxies, analysis of the stellar populations of galaxies with over- and under-massive black holes has been used to argue for AGN feedback directly influencing star formation histories (e.g., \citealt{2018ApJ...855L..20M}). Searching for similar trends between galaxy properties and position on the $\mathrm{M_{BH}-\sigma_*}$ relation will further inform our picture of AGN feedback in low mass galaxies. Further quantifying the scatter in the relationship is also key, as \cite{2017ApJ...839L..13S} suggests that dwarf galaxies following $M_{\rm BH}-\sigma_{\ast}$ while falling below the $M_{\rm BH}-M_{\ast}$ relation could be evidence for AGN-driven suppression of star formation. \\

In summary, we have measured stellar velocity dispersions for eight active dwarf galaxies, doubling the number of dwarf galaxies on the $M_{\rm BH}-\sigma_{\ast}$ relation. These galaxies have BHs with masses ranging from $10^{5}-10^{6}\;M_{\odot}$. We find that these systems are in good agreement with the extrapolation of the $M_{\rm BH}-\sigma_{\ast}$ relation, implying that the processes which lead to the present-day $M_{\rm BH}-\sigma_{\ast}$ relation also apply to dwarf galaxies. This provides important constraints for models of BH formation and growth. Larger samples of dwarf galaxies with measured stellar velocity dispersions and BH masses are necessary to determine which properties impact BH growth in these systems. 

\facility{Keck:II (ESI)}

\acknowledgements
The authors thank the anonymous referee for their comments and suggestions which have improved this manuscript. Support for VFB was provided by the National Aeronautics and Space Administration through Einstein Postdoctoral Fellowship Award Number PF7-180161 issued by the Chandra X-ray Observatory Center, which is operated by the Smithsonian Astrophysical Observatory for and on behalf of the National Aeronautics Space Administration under contract NAS8-03060. 

The data presented herein were obtained at the W. M. Keck Observatory, which is operated as a scientific partnership among the California Institute of Technology, the University of California and the National Aeronautics and Space Administration. The Observatory was made possible by the generous financial support of the W. M. Keck Foundation. The authors wish to recognize and acknowledge the very significant cultural role and reverence that the summit of Maunakea has always had within the indigenous Hawaiian community.  We are most fortunate to have the opportunity to conduct observations from this mountain.

This research has made use of the NASA/IPAC Extragalactic Database (NED) which is operated by the Jet Propulsion Laboratory, California Institute of Technology, under contract with the National Aeronautics and Space Administration.

\bibliographystyle{apj}

\clearpage

\begin{figure*}
\centering
\includegraphics[width=0.42\textwidth]{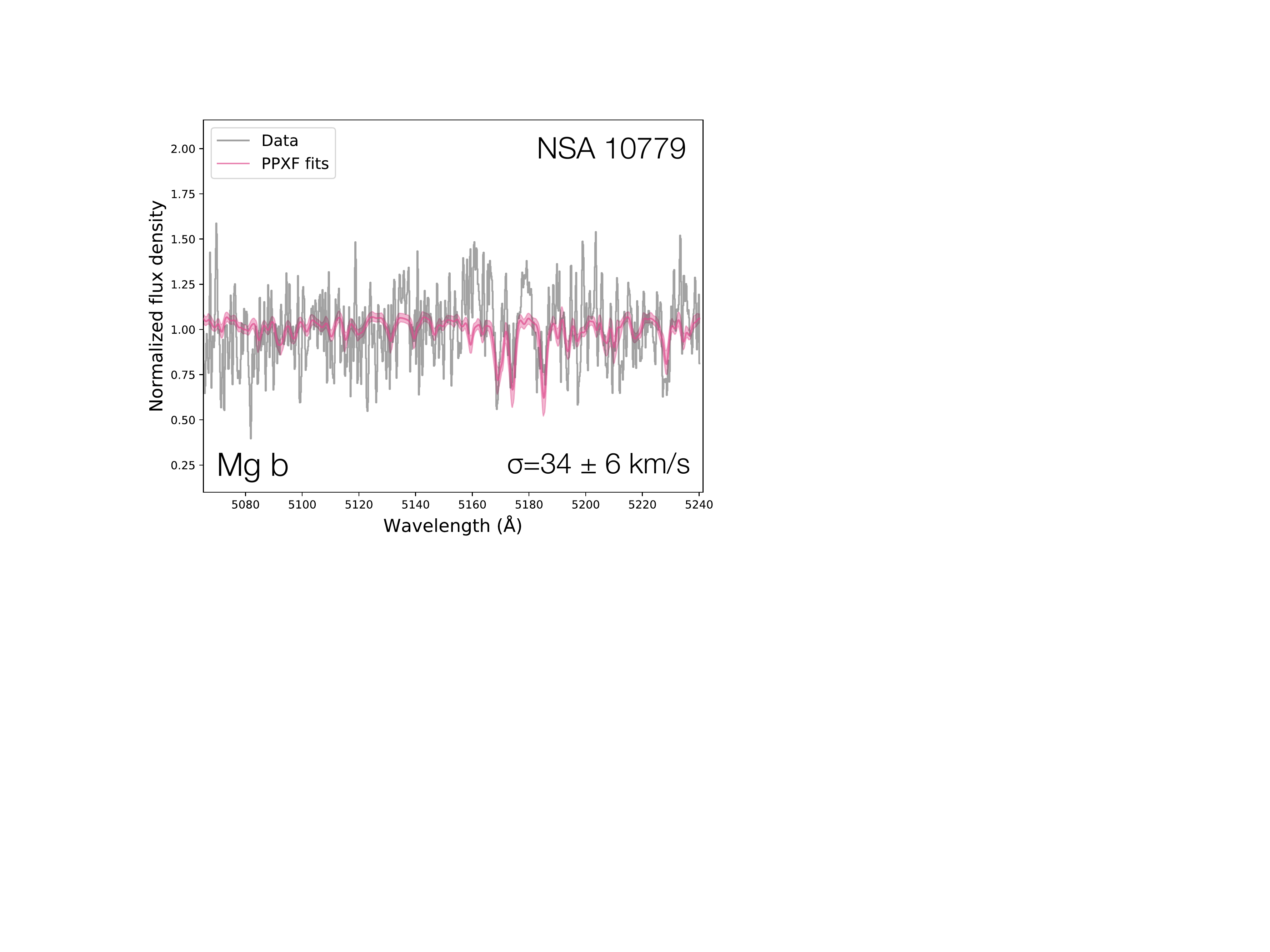}
\includegraphics[width=0.84\textwidth]{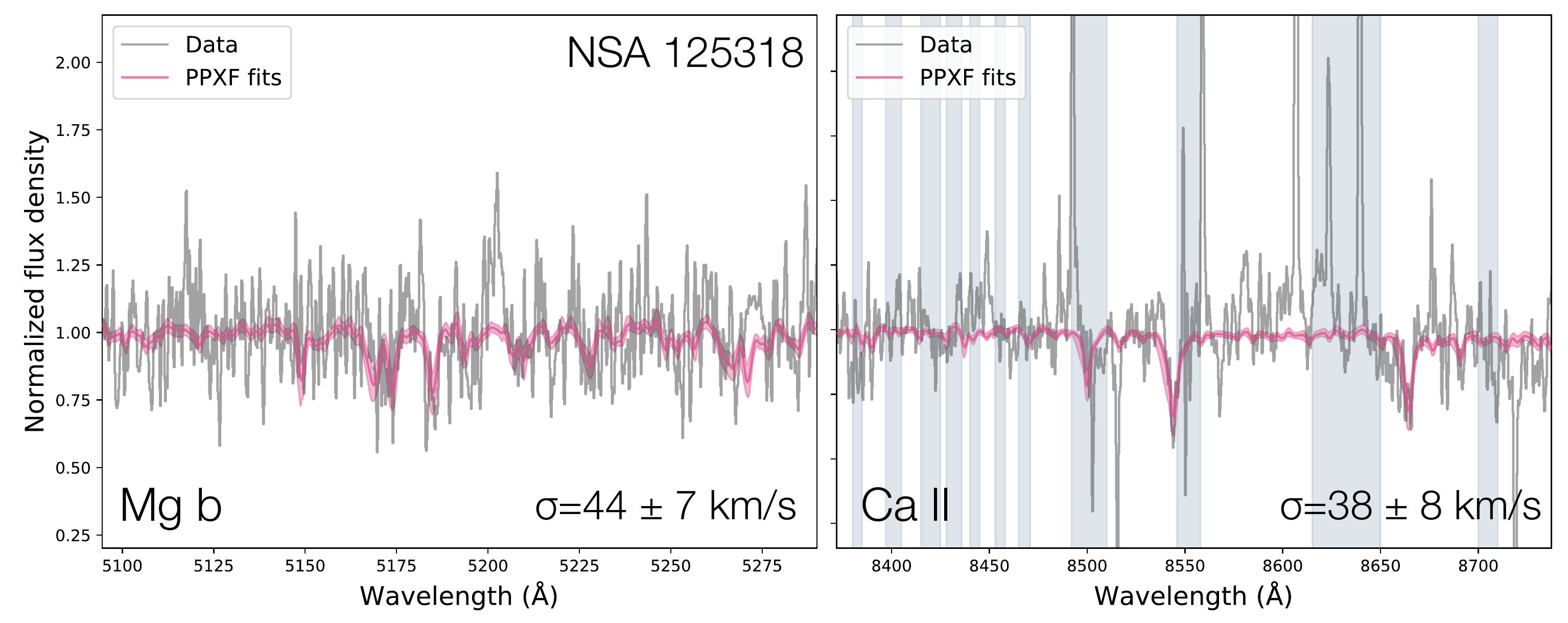}
\includegraphics[width=0.84\textwidth]{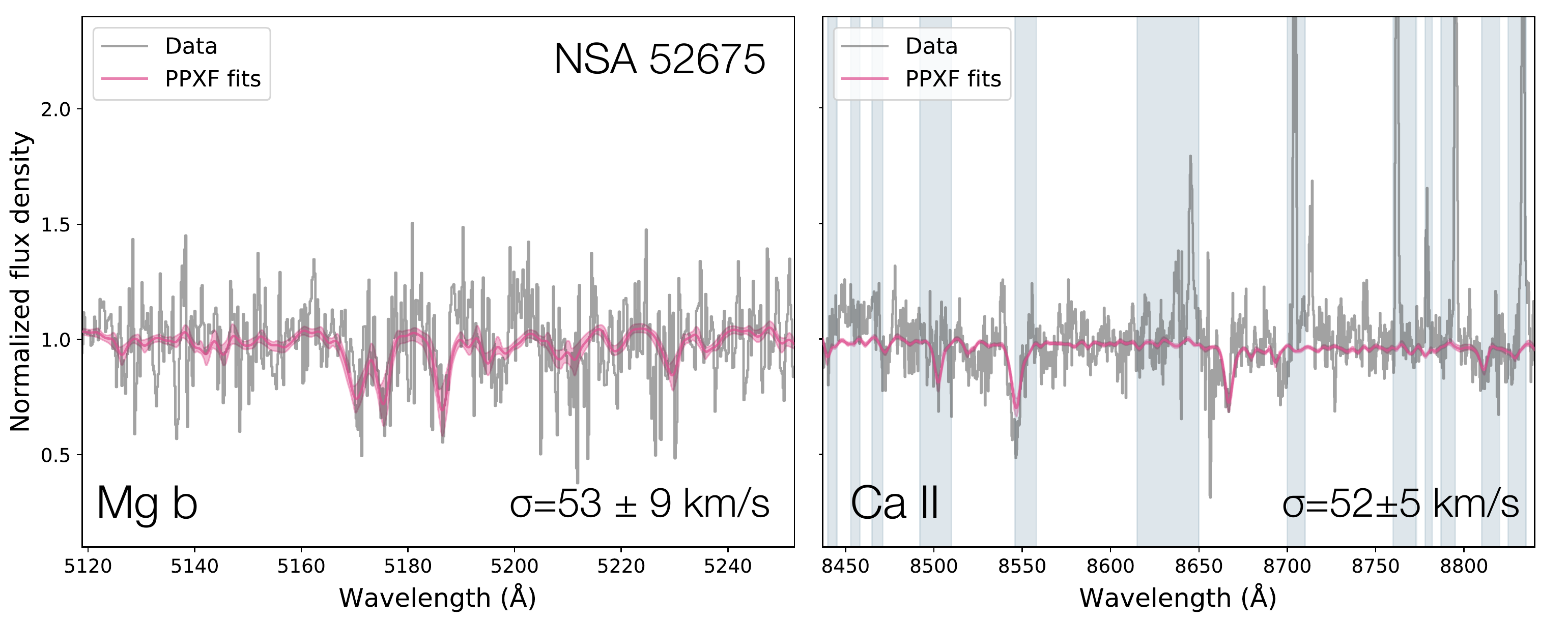}
\caption{Same as Figure~\ref{nsa52675_panels}, but for additional objects. Spectra are smoothed with a box size of 3 pixels for plotting. }
\label{additional_sigs}
\end{figure*}

\begin{figure*}
\centering
\includegraphics[width=0.42\textwidth]{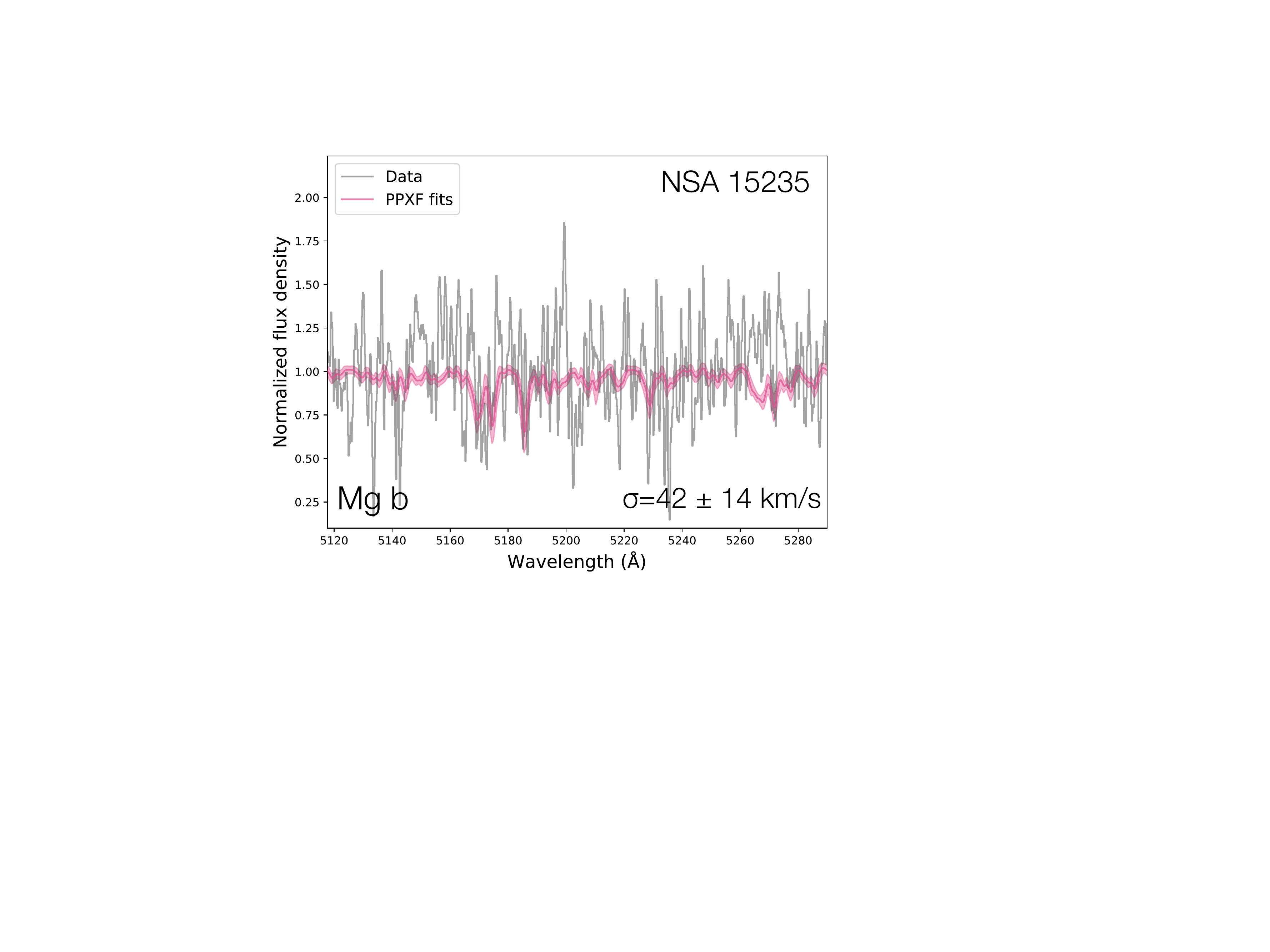}\\
\includegraphics[width=0.42\textwidth]{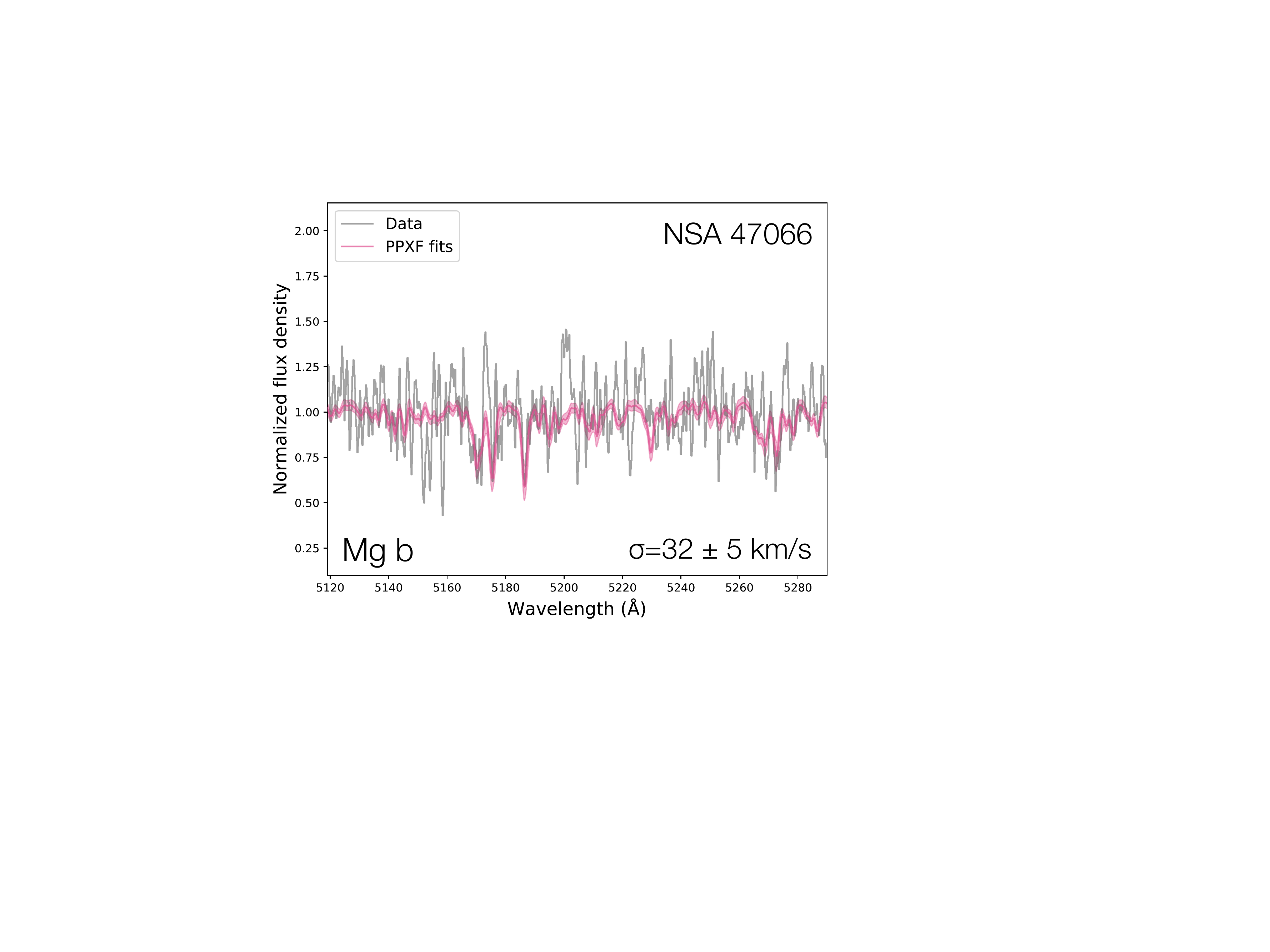}\\
\includegraphics[width=0.42\textwidth]{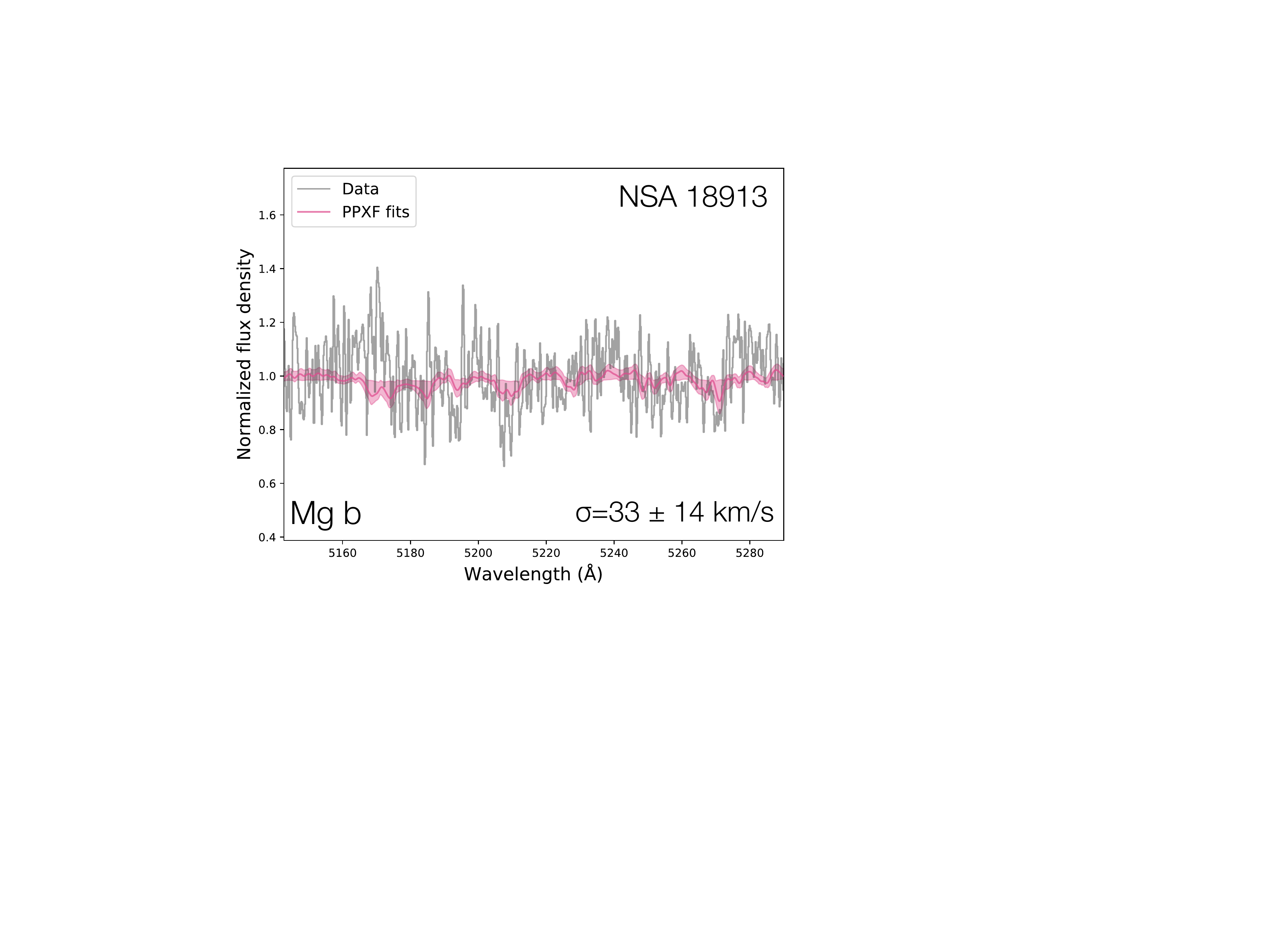}\\
\includegraphics[width=0.84\textwidth]{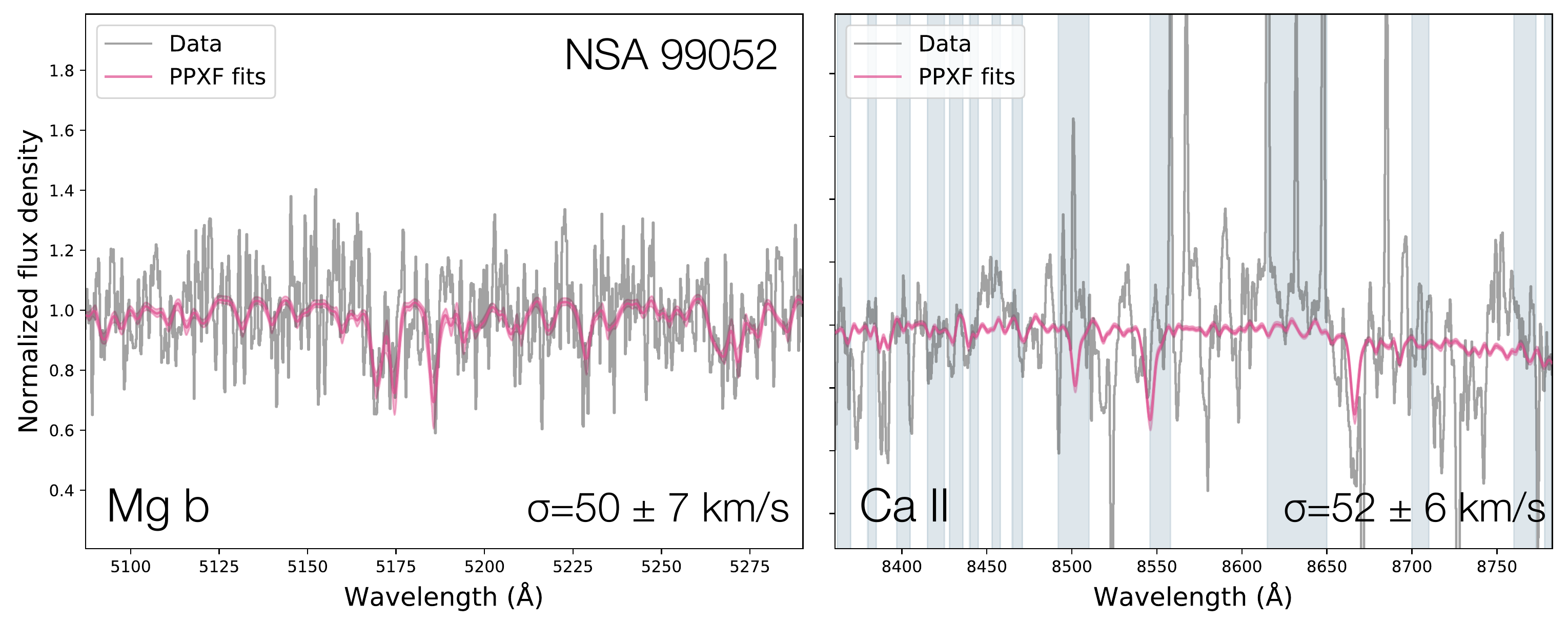}
\caption{Same as Figure~\ref{nsa52675_panels}, but for additional objects. Spectra are smoothed with a box size of 3 pixels for plotting.}
\label{additional_sigs2}
\end{figure*}

\end{document}